\documentstyle[preprint,aps]{revtex}

\def\k{k}

\begin{document} \draft
\begin{flushright}UF-IFT-HEP-97-7\\ August 1997
\\Revised: April 1998\end{flushright}
\centerline{\large \bf MINIMUM ENTROPY PRODUCTION OF NEUTRINO RADIATION}
\centerline{\large \bf IN THE STEADY STATE}
 \vskip 0.6cm
\centerline{\bf Christopher Essex}
 \vskip 0.6cm
\centerline{Department of Applied Mathematics}
\centerline{The University of Western Ontario, London, Canada  N6A 5B7}
\vskip 0.6cm
\centerline{\bf Dallas C. Kennedy}
 \vskip 0.6cm
\centerline{Department of Physics}
\centerline{University of Florida, Gainesville FL 32611 USA}
\vfill \eject

\pagestyle{plain}
\begin{abstract}
A thermodynamical minimum principle valid for photon radiation 
is shown to hold for arbitrary geometries. It is successfully extended to
neutrinos, in the zero mass and chemical potential case, following a
parallel development of photon and neutrino statistics. This
minimum principle stems more from that of Planck than that of classical
Onsager-Prigogine
irreversible thermodynamics. Its extension from bosons to fermions
suggests that it may have a still wider validity.
\end{abstract}

\vfill \eject

\section{Introduction}

Photon radiation has a distinctive quality in that it interacts little enough 
with matter that it is typically far from thermal equilibrium
even when matter may be close to equilibrium. This is the origin of classical
radiative transfer
\cite{Sch,Ch}, which represents a link
between
far-from-equilibrium properties of radiation to the near-equilibrium 
thermodynamics of matter, leading to some curious thermodynamical consequences
\cite{Essexa,Essexb}.  These in particular concern minimum entropy
production
in a non-equilibrium steady state (NESS).  This is a property distinct from
the classical theory of
irreversible thermodynamics \cite{DeG}, as few of the
requirements for that theory hold: one does not even have a local 
thermodynamical bilinear form of entropy production
\cite{Essexa,Essexad}.

Is this distinctive linkage one that may only be found in the domain of
bosons? The
answer is surely not, as neutrinos, which are fermions, are even more tenuously
linked to matter than photons. However, a beam of neutrinos, unlike one 
of photons, is not at all the norm. In the case of fermions, in
contrast,
we must turn to the atypical domains of a supernova and the early Universe to 
have macroscopic beams interacting meaningfully at a thermodynamical level 
with matter.  These physical situations involve a boundary in space or time
between neutrino equilibrium and non-equilibrium.
If we allow neutrino production without 
effective absorption, ordinary stars provide another example domain.

However, the question is to what extent the thermodynamical properties 
that hold for photons also hold for neutrinos. That is, is there a similar 
minimum principle for entropy production in the case of neutrino radiation? 
This paper addresses this question by generalizing ``radiation'' to include
any exactly or nearly massless particles whose number is not conserved.
We consider only the case of matter in local thermal equilibrium (LTE) and 
where macroscopic thermodynamics is valid, not the more general case 
when LTE breaks down and kinetic theory is necessary \cite{BoerGroot}.

We proceed here with the assumption that the rest mass and chemical
potential of neutrinos are zero. Clearly a more comprehensive treatment must
include the possibility that neither is zero \cite{EaK}.
However there are advantages to proceeding in this manner in the first 
instance. By making this choice we ensure that the neutrinos are as much like 
the photons as possible.  The rest mass is not so much a
problem here
as is the chemical potential.  A non-zero chemical potential presents a 
qualitative departure from the thermodynamic properties of photons in that 
it implies an additional independent thermodynamical variable. However, what 
is learned in this case can be a guide to future work.

We find that neutrinos have an entropy production minimum principle in 
the steady state similar to that of photons, which also manifests 
itself as a conservation principle for energy. Implicit in weak
reactions involving neutrinos is the conservation, not only of electric
charge $Q,$ but of lepton number $L$ and baryon number $B$ as
well. $Q,$
$B,$ and $L$ conservation in weak reactions play a non-trivial role, unlike
$Q$ in purely electromagnetic processes.  These quantities are assumed to be
exactly conserved in the microscopic sense. But conservation in this
paper may also have a secondary thermodynamical meaning:
absence of sources or sinks of these numbers in the form of
macroscopic, thermodynamic reservoirs.

We proceed in a parallel manner between photons and neutrinos in order to 
highlight differences and similarities and to link with the previous work on 
photons.

\section{General Definitions and Relations}

Phase space for both neutrinos and photons is defined by position ${\bf r},$
momentum ${\bf p},$ implying energy $\epsilon $. 
In the case
of photons it
is often customary to introduce frequency $\nu$ as a proxy for energy and 
wavenumber ${\bf k}$ instead of momentum.  The energy of unpolarised
particles in the phase space volume $d^3p~d^3r$ is
\begin{equation}
2 n \epsilon \frac{d^3p~d^3r}{h^3}\quad ,
\end{equation}
where $n$ is the mean occupation number for either photons or
neutrinos. $h$ is Planck's
constant. The entropy of that same volume is \begin{equation}
2 \k [ \mp (1 \mp n) \ln (1\mp n) - n \ln n] 
\frac{d^3p~d^3r}{h^3}\quad ,
\end{equation}
where $\k$ is Boltzmann's constant. The upper signs correspond to 
neutrinos 
and the lower to photons.

A momentum-dependent temperature $T_{\bf p}$ may be introduced for 
this small 
phase volume, by forming the derivative of entropy with respect to 
energy,
\begin{equation}
\frac{1}{T_{\bf p}} = \frac{\k}{\epsilon}\ln (n^{-1}\mp 1)\quad ,
\end{equation}
which takes its physical meaning from the time-independent steady state of
noninteracting quanta. This makes the cell indistinguishable from one
that shares the same temperature with all other cells. We extend this naturally
to neutrinos given the assumptions of the paper: the rest mass $m_\nu$ and
the chemical potential $\mu_\nu$ are both zero.  One finds
\begin{equation}
\frac{d^3p~d^3r}{h^3} = 
\left(\frac{\k T_{\bf p}}{hc}\right)^3~x^2~dx~d\Omega~d^3r\quad ,
\end{equation}
given the new variable $x$ such that $\epsilon$ = $pc,$ 
$x=\epsilon/{\k T_{\bf p}}$, and $d^3p$ = $p^2 dp 
d\Omega$, where $d\Omega$ is an element of solid angle.

In thermal equilibrium,
\begin{equation}
n_{\bf p} =  \frac{1}{e^x \pm 1}\quad ,
\end{equation}
and we may drop the subscript ${\bf p}$ on the temperature.
The energy in the phase volume may then be written
\begin{equation}
\frac{2(\k T)^4}{(hc) ^3} \frac{x^3}{(e^x \pm 1)}~dx~d\Omega~d^3r
\quad .
\end{equation}
From this we note that when an integration over $x$ is carried out that the 
fourth-power law follows for both photons {\it and} neutrinos. The only 
difference between the two is in the numerical factor of the integral due to 
the ``$\pm$'' in the denominator in the integrand of the $x$ integration.
The entropy of the phase volume is treated similarly. After some 
simple manipulations it becomes
\begin{equation}
\frac{2\k (\k T)^3}{(hc)^3}\Big[ \frac{x^3}{(e^x \pm 1)} \pm x^2 \ln (1
\pm e^{-x}) \Big] dx~d\Omega~d^3r\quad .
\end{equation}
This implies the expected third-power behaviour for entropy in equilibrium, 
for neutrinos as well as photons.

The four integrals:
\begin{equation}
\int^\infty_0\ \frac{x^3}{(e^x \pm 1)} \ dx = \frac{15 \mp 1}{16}
\Big(\frac{\pi^4}{15}\Big)\quad ,
\end{equation}
and
\begin{equation}
\int^\infty_0\ \pm x^2 \ln (1 \pm e^{-x}) \ dx = \frac{15 \mp 1}{16}
\Big(\frac{1}{3}\Big)\Big(\frac{\pi^4}{15}\Big)\quad ,
\end{equation}
are easily deduced by simple series expansions.
From these we find the energy per unit volume into solid angle
$d\Omega$,
\begin{equation}
\frac{15 \mp 1}{16}\Big(\frac{\sigma}{\pi c}\Big)T^4d\Omega\quad,
\end{equation}
where $\sigma$ = $2\k^4\pi^5/(15c^2h^3),$ the Stefan-Boltzmann constant.  
Similarly for the entropy,
\begin{equation}
\Big(\frac{4}{3}\Big)\frac{15 \mp 1}{16}\Big(\frac{\sigma}{\pi c}\Big)T^3
d\Omega\quad .
\end{equation}
The vector flux density of energy into solid angle $d\Omega$ with 
direction $\hat{\bf m}$ is
\begin{equation}
\frac{15 \mp 1}{16}\Big(\frac{\sigma}{\pi}\Big)T^4\hat{\bf m}~d\Omega
\quad ,
\end{equation}
and for entropy,
\begin{equation}
\Big(\frac{4}{3}\Big)\frac{15 \mp 1}{16}\Big(\frac{\sigma}{\pi}\Big)T^3 
\hat{\bf m}~d\Omega\quad .
\end{equation}
The integrals~(8,9) give the canonical fermion factor of $7/8$ relative to
bosons.  We conclude that the flux density per solid angle, known 
variously 
as the specific intensity or radiance, is for energy,
\begin{equation}
\frac{15 \mp 1}{16}\Big(\frac{\sigma}{\pi}\Big)T^4\quad ,
\end{equation}
and for entropy,
\begin{equation}
\Big(\frac{4}{3}\Big)\frac{15 \mp 1}{16}\Big(\frac{\sigma}{\pi}\Big)T^3\quad .
\end{equation}

Even out of equilibrium, we may relate the specific intensity
$I_\epsilon$, for a given $\epsilon$, to $n$ from equations (1) and (4) 
and the {\it flux density} into a solid angle,
\begin{equation} I_\epsilon =
 \frac{2n_\epsilon\epsilon^3}{h^3c^2}\quad .
\end{equation} 
The specific entropy intensity is 
\begin{equation}
J_\epsilon  =  \frac{2\k\epsilon^2}{h^3c^2}\Big[ \mp \Big(1 \mp 
\frac{c^2h^3I_\epsilon}{2\epsilon^3}\Big)
\ln\Big(1\mp \frac{c^2h^3 I_\epsilon}{2\epsilon^3}\Big) \nonumber \\ 
 - \Big(\frac{c^2h^3 I_\epsilon}{2\epsilon^3}\Big)
\ln\Big(\frac{c^2h^3 I_\epsilon}{2\epsilon^3}\Big)\Big]\quad .
\end{equation}
The fundamental extensive quantities are the specific entropy flux 
$J_\epsilon$ and specific energy flux $I_\epsilon$. Note that expression
(3) is recovered by forming $dJ_\epsilon /dI_\epsilon$.

Although neutrino number is not conserved, lepton number is, and it is
thus physically important to define a specific {\it number}
flux for neutrinos $N_\epsilon$ corresponding to $I_\epsilon$:
\begin{equation}
N_\epsilon = \frac{2n_{\bf p}\epsilon^2_{\bf p}}{h^3c^2}\quad ,
\end{equation}
where
\begin{equation}
I_\epsilon = \epsilon N_\epsilon\quad .
\end{equation}

In the case of photons, number is not interesting, as photon number is not
conserved.  In the case of fermions, some number conservation law always
holds (because of the half-integral spins).  With zero chemical potential,
however, the neutrino number is a purely auxiliary quantity and depends on
the energy flux.  Conversely, we could take neutrino number as fundamental
and energy as derived; in either case, only one variable is
independent.
In the case of non-zero chemical potential, not treated in this paper,
neutrino number and energy flux become independent variables, with an
energy-dependent chemical potential $\mu_\epsilon$.

\section{Entropy Production and Entropy Flows}

Entropy is an extensive thermodynamic property and can be localised and 
integrated to determine a global amount.  Localization is also possible for 
entropy production itself.  This result agrees with the general principle of
the locality of physical interactions, so long as a thermodynamical picture is
valid.  It also means that thermodynamics is not restricted to a macroscopic 
box. If we divide space into distinct regions, boundary surfaces are defined;
entropy can be moved between regions and the notion of entropy flux across a
surface follows.

The entropy production rate can be expressed in a balance or conservation 
equation,
\begin{equation}
\epsilon_s = {{\partial s} \over {\partial t}} + {\bf\nabla\cdot\bf{\cal F}}~,
\end{equation}
where $\epsilon_s $ is the entropy production rate per unit volume, $s$ is the
volume density of entropy, and  ${\bf\cal F}$ is the entropy flux 
density.

It is only by 
convention that this equation is called a {\it conservation} equation.  It is 
really an accounting equation, with no implication of conservation.  In 
fact, this equation allows a general statement of
the second law of thermodynamics: $\epsilon_s \geq 0$.

A somewhat artificial distinction may be made now in equation~(20), between
radiative and matter processes.  In the former case we necessarily must
always consider full phase space, while in the latter we assume a 
near-equilibrium distribution in energy, nearly identical in all directions 
(so-called ``local thermal equilibrium'' or LTE).  In that case we find
\begin{equation}
\epsilon_s = \epsilon_s^m + \epsilon_s^\gamma~,
\end{equation} 
where the superscript $m$ denotes non-radiative components which shall
be  termed ``matter."  $\gamma$ denotes radiative components.
Writing these entropy sources out explicitly we have,
\begin{equation}
\label{Bal}
\epsilon_s = {{\partial
s_m} \over {\partial t}}+ {{\partial s_\gamma } \over {\partial t}} +
\nabla \cdot {\bf Y}_s + \nabla \cdot {\bf H}~,
\end{equation} 
where $s_m$ is the volume density of
entropy
in matter, and $s_\gamma$ is the volume density of entropy in radiation.
${\bf Y}_s$ and {\bf H} denote non-radiative and
radiative entropy flux densities, respectively.

As radiation we mean here, of course, photons or neutrinos, while other
particles
take the role of being non-radiative and in LTE.  A mixture of 
near-equilibrium (LTE) matter and the far-from-equilibrium radiation is a
typical one in the Universe.  It is the mixture in which you are
immersed while reading this page: you are warm, and yet you can read
these words radiatively.

By using the equations of state and balance equations for extensive 
variables, we re-express the entropy production rate in the
form \cite{Essexg}
\begin{equation}
\epsilon_s = \sum_{k} \{ a_k \epsilon_k + \nabla a_k \cdot {\bf Y}'_k \}
+ {{\partial s_r } \over {\partial t}} + \nabla \cdot {\bf H}~,
 \end{equation} 
where the sum is over contributions from extensive variables with index $k$. 
$a_k$ is the conjugate intensive variable divided by the temperature.
$\epsilon_k$ is the creation rate of variable $k$ (for example, the rate
that internal energy is created from the radiation field, nuclear
reactions, or viscous dissipation).
${\bf Y}'_k$ is the flux density of variable $k$ (for example, the flux 
density of internal energy in the case of diffusion).
The prime denotes the value in the rest frame of the medium.
As massless radiation is without a rest frame, the separation of the
entropy production into these two parts thus turns out to be not at all 
artificial.

If we assume in~(\ref{Bal}) a steady radiation field, and integrate over a 
finite volume $V$ bounded by a surface $S,$ with element $d{\bf S},$
containing all of the matter, then the overall entropy production rate
$\Sigma$ is
\begin{equation}
\Sigma = \int_V {{\partial
s_m} \over {\partial t}} dV + \int_S {\bf H}\cdot d{\bf S}~,
\end{equation} 
because matter fluxes must vanish across $S$.

If we ignore matter transport processes, equation~(23) becomes
\begin{equation}
\label{Glob}
\Sigma = \int_V \sum_{k} \{ a_k \epsilon_k \}\ dV + \int_S {\bf H}\cdot
d{\bf S}~.
\end{equation}
We may arrive at this result, alternatively, by imagining that the process 
is steady and that the entropy change of matter (the first integral) is 
reversibly drained off to a heat or another type of reservoir. That is, for 
this steady case,
we deal only with a subsystem, thermodynamically speaking, and so
conservation laws  do {\it not} hold {\it macroscopically} 
(i.e. integrated) in the
subsystem alone. Of course this has no bearing on {\it microscopic}
conservation laws.

It is worth noting that the terms under the first integral of
(\ref{Glob}) are all due
to processes in matter and are not a part of the entropy production for photons
or neutrinos.  It is a common misconception to interpret the radiation heating
rate over the temperature, which is a possible term under the first integral, 
as the entropy production of radiation.  It should be clear from this 
construction that $\int_V \epsilon_s^\gamma dV$ is all accounted for
through the second integral of (\ref{Glob}).

\section{Minimum Entropy Production}

Equation (\ref{Glob}) provides a structure for computing the
entropy
production rate due to the interaction of matter and radiation for 
many
finite bodies locally in equilibrium. The first term represents whatever 
changes are manifest in the entropy of the body while the second term 
accounts for changes in the radiation field itself due to the 
interaction.

If photon radiation is impinging on a body of temperature $T$
in a vacuum, the entropy production rate is from (25)  just
\begin{equation}
\Sigma = \int_V \left(-\frac{\nabla \cdot {\bf F}}{T}\right) dV + \int_S {\bf 
H}\cdot d{\bf S}~,
\end{equation}
where the volume $V$ is any containing the body, and ${\bf F}$ is the 
volume flux density of energy radiation. The latter is the integral over
all solid angles and energy of $I_\epsilon {\bf \hat{n}}$, where ${\bf
\hat{n}}$ is the unit vector defining the direction that $I_\epsilon$
(16) flows in. If the temperature
is (artificially held) uniform over the body, then \begin{equation}
\Sigma = - \frac{1}{T} \int_S {\bf F} \cdot d{\bf S} + \int_S
{\bf H}\cdot d{\bf S}~.
\end{equation}
If the surface area of the body is $A$, and it emits as a black body, 
then
\begin{equation}
\Sigma = \frac{1}{T} \left\{ \left| \int_S {\bf F}^i \cdot d{\bf S} \right| -
\sigma
T^4 A \right\} - \left| \int_S {\bf H}^i \cdot d{\bf S} \right| + \frac{4}{3}
\sigma T^3 A~, \end{equation}
where the remaining integrals represent impinging photon radiation, and 
the superscript $i$ denotes an impinging flow only, which is 
independent of the state of the body.

It follows that
\begin{equation}
\frac{d\Sigma}{dT} = -\frac{1}{T^2} \left\{ \left| \int_S {\bf F}^i \cdot
d{\bf S} \right| -  \sigma T^4 A \right\} ~, \end{equation}
or for a minimum
\begin{equation}
 \left\{ \left| \int_S {\bf F}^i \cdot d{\bf S} \right| -  \sigma T^4 A
\right\} = 0~.
\end{equation}
That is, the entropy production rate is a minimum in the  steady
state,  implying energy conservation \cite{Essexa,Essexc}, but for
for an arbitrary geometry and impinging field. 

Equations (26) to (30) could represent the entropy production of a (black)
planet irradiated by photons originating elsewhere from a star,
or could represent a gas cloud under similar photon radiation.  They could
also represent the entropy production of a target of matter
irradiated by a laser.   
One can envisage matter in the laboratory as
being connected to a heat bath, or circumstances where the heat capacity
is very high, to justify the steady, isothermal construction of the rate
of entropy change in matter (first term in (26)).  Obviously the
isothermal construction may be relaxed \cite{Essexc} using the equations
of Section III, but that is beyond the scope of this paper.

While these equations represent an extension of
previous work \cite{Essexa,Essexc} to arbitrary matter and radiation 
geometries, the remarkable thing about them is that they represent a 
minimum entropy production principle that is foreign to the 
classical minimum entropy production principle
of Prigogine \cite{DeG}. There are no Onsager reciprocity relations, no
``linear" empirical flux-force laws, and no meaningful 
thermodynamical fluxes and forces. In the latter regard,
there need only be one temperature defined for the 
entire problem.

The question of whether this remarkable property is restricted to boson
radiation in the from of photons has not been put previously, so we 
now consider the corresponding problem for
fermions in the form of neutrinos. As in the case of photons we
turn to equation (\ref{Glob}), but at this point the differences between
photons and neutrinos emerge, not in the second (radiation) term,
but in the first (matter) term. That is because of the 
different manner in which neutrinos interact with matter. While
photons do not conserve their number and trivially conserve electric
charge, neutrinos are linked to the conservation of charge, lepton number, and
baryon number through the structure of weak interactions. Here we consider 
only first-generation fermions and only nucleons for hadrons:
 \begin{equation}
\label{Reac}
 \nu + n  \rightarrow p + e^- ~,
\end{equation}
and all related reactions.
Thus for neutrinos (\ref{Glob}) becomes
\begin{equation}
\Sigma = \int_V \left(-\frac{\nabla \cdot {\bf F}}{T} + \frac{\mu_e
\dot{n}_e + \mu_n
\dot{n}_n +\mu_p
\dot{n}_p}{T} \right) dV  + \int_S {\bf H}\cdot d{\bf S}~.
\end{equation}
Recall that $\mu_\nu = 0$ is assumed for neutrinos.
$\dot{n}_e$, $\dot{n}_n$, and $\dot{n}_p$ represent rates of 
change of number densities for electrons, neutrons and protons
respectively, each of which is multiplied by its corresponding chemical 
potential. Assuming chemical equilibrium within the matter,
\begin{equation}
\mu_e
  +\mu_p - \mu_n
 = 0   ~.
\end{equation}
This, together with an isothermal and black body (neutrino) assumption,
leads to
\begin{equation}
\Sigma = \frac{1}{T} \left\{ \left| \int_S {\bf F}^i \cdot d{\bf S} \right| -
\frac{7}{8}\sigma
T^4 A \right\} + \int_V \frac{1}{T} \left\{  \mu_n (\dot{n}_n +\dot{n}_p) + 
\mu_e (\dot{n}_e  - \dot{n}_p) \right\} dV- \left| \int_S {\bf H}^i \cdot 
d{\bf S} \right| + \frac{7}{8}\cdot \frac{4}{3} \sigma T^3 A~.
\end{equation}

Conservation of baryon number and charge produce
\begin{equation}
\Sigma = \frac{1}{T} \left\{ \left| \int_S {\bf F}^i \cdot d{\bf S} \right| -
\frac{7}{8}\sigma
T^4 A \right\} - \left| \int_S {\bf H}^i \cdot d{\bf S} \right| +
\frac{7}{8}\cdot \frac{4}{3} \sigma T^3 A~.
\end{equation}

Except for the factors of $\frac{7}{8}$, this equation is identical to 
(28). Thus in minimum entropy production, the energy balance steady 
state \begin{equation}
 \left\{ \left| \int_S {\bf F}^i \cdot d{\bf S} \right|-  \frac{7}{8} \sigma
T^4 A \right\} = 0 ,
\end{equation}
must hold as well. Thus we find that this distinctive minimum entropy
production result is extended to neutrinos, and thereby beyond the 
limitation to bosons to at least some fermions.

\section{Local and Nonlocal Regimes}

Generally, the interactions of neutrinos and photons with matter are most
simply viewed as purely local.  For statistical physics, however, we also
need to count momentum states.  If we make the matter-radiation separation
of section~3 and further assume LTE for matter, the matter momentum states
can be integrated out, leaving the full phase space only for radiation.  This
condition allows us to avoid a full kinetic theory calculation 
\cite{BoerGroot}.
At this juncture, we have a choice of local versus action-at-a-distance 
representation for the radiation.

While the radiation exists in its own right, in the event that the
overall entropy of the radiation field is not changing, we need only be
concerned with matter-matter interactions mediated by radiation, as the
radiation terms can be then be integrated out.  Radiation then becomes
merely a special kind of nonlocal heat transport, and $\Sigma$ may be
represented in a multilocal form.  This multilocal form is at least true
for the first term in (\ref{Glob}) even when separate changes do take
place in the radiation field, such as in conservative scattering processes 
\cite{Essexad}.  In all cases, it is the {\it total} $\Sigma$ that is
minimised.

In the case of quanta in an extended, continuous medium, a common
matter-radiation LTE is often valid, with a common matter-radiation
temperature $T({\bf r})$. This temperature in general is not constant
in space.
LTE holds to extremely high accuracy for {\em photons} inside the
photospheric surface of a star, for example, but not for neutrinos.  
The entropy production associated with the
production and transport (diffusion) of photons is
\begin{eqnarray}
\label{Froin}
\Sigma_\gamma & = &
\int~dV~\Big\{ (1/2)[4acT^5/(3\kappa_\gamma )]
  [{\bf\nabla}(1/T)]^2
 + \varepsilon_\gamma /T\Big\}~,
\end{eqnarray}
where $\varepsilon_\gamma$ is the photon energy production rate density
and $\kappa_\gamma$ is the opacity (inverse mean free path) of the
matter against photon diffusion \cite{KaB}.
Equation (\ref{Froin})  corresponds to 
the second term in equation (\ref{Glob}), but written as a volume 
integral of a divergence, up to but {\em not} including the photosphere.
This is in contrast to photon entropy production
discussed previously in this paper, in that the photons here are virtually 
in equilibrium with matter and so are diffusive, not radiative.

At the photospheric surface, a single
LTE ceases to hold (see below), and the diffusive approximation of
equation (\ref{Froin}) breaks down. Nonetheless the second term of 
(\ref{Glob}) still represents the entropy production in the radiation
field, but at the photosphere and outside, the photons become 
radiative. The complete photon entropy
production of a star of radius $R$ (including
its photosphere) is
 \begin{equation}
\label{all}
\Sigma_{\gamma ,\rm sur} = \frac{4}{3}\sigma T^3_{\rm sur}(4\pi R^2)
\end{equation}
for the photon radiation released into empty space at an idealised sharp 
surface.  $(T_{\rm sur}$ is
the photospheric surface temperature.)    As the volume 
of integration is increased $\Sigma_\gamma$ picks up additional
contributions  to interactions with more matter, for
example, with a planet \cite{Essexa,Essexb,Essexc,Les}.

Neutrinos emitted by ordinary stars are quite different from photons: as the
interior temperatures are not high enough for weak interactions to be in
LTE, the neutrinos are not emitted in anything like a
blackbody distribution, and are not subsequently thermalised.  Their
spectra are instead determined almost exactly
by the microscopic reaction spectra and emerge essentially unaffected by
the neutrinos' subsequent travel through stellar matter to empty space.
If the emitting star does not absorb neutrinos and the receiving 
Earth-bound detector does not emit neutrinos, the total neutrino entropy 
production is
 \begin{eqnarray}
\Sigma_\nu & = & \int_{\rm emitter}~dV~\int~d^3p\ 
\frac{\epsilon_{\bf p}\ {\dot n}_{\bf p}}{T_{\bf p}} - \nonumber \\
 & & \int_{\rm receiver}~dV^\prime~\int~d^3p^\prime\
{\epsilon_{\bf p^\prime}\ {\dot n}_{\bf p^\prime}\over T_{\bf p^\prime}}~,
\end{eqnarray}
where ${\dot n}_{\bf p}$ $({\dot n}_{\bf p^\prime})$ is the neutrino production
(absorption) rate density in real
and momentum space.  In a NESS, ${\dot n}_{\bf p}$ depends only
on
$\epsilon_{\bf p},$ not on emission direction ${\bf\hat p}.$  

In an equilibrated supernova or the early Universe, on the other hand, the 
neutrinos 
are emitted and absorbed in LTE.  The entropy production below the supernova
neutrinosphere is a function of a single local temperature:
\begin{eqnarray}
\Sigma_\nu & = &
\int~dV~\Big\{ (1/2)[7acT^5/(6\kappa_\nu )]
[{\bf\nabla}(1/T)]^2
 + \varepsilon_\nu /T\Big\}\quad ,
\end{eqnarray}
like~(\ref{Froin}), with a neutrino mean free path $1/\kappa_\nu$ and
an extra factor
of $7/8$ in the diffusion part. 
The total $\Sigma_\nu$ {\it including} the neutrinosphere is analogous 
to (\ref{all}).  An analogous expression can be constructed for photon and
neutrino entropy production in the early Universe before their respective
decouplings, but this would require inclusion of general relativity.

Even if the neutrinos or photons are emitted and absorbed locally as a gas, 
the system in general is not in equilibrium with a single temperature $T$
or $T({\bf r}).$  Multiple temperatures can be defined if each system 
component retains its own LTE, with each component spectrum thermal.
For example, a photon gas with a frequency-dependent
temperature $T_\gamma$ may interact with matter of temperature $T.$  Then
\begin{eqnarray}
\Sigma_\gamma = \int dV \int d\epsilon \int d\Omega\ 
I_\epsilon ({\bf r},\epsilon ,\Omega )
 \Bigl[\frac{1}{T_\gamma({\bf r},\epsilon ,\Omega )}
- \frac{1}{T({\bf r})}\Bigr]\quad .
\end{eqnarray}
$I_\epsilon$ is the local specific energy intensity of photons
emitted {\it by} the matter.  If $T_\gamma >$ $T,$ then $I_\epsilon\le$
0; if $T_\gamma <$ $T,$ then $I_\epsilon\ge$ 0.  Thus $\Sigma_\gamma$ is always
$\ge$ 0.  Stellar atmospheres provide a related example.  The radiation has a 
temperature
$T_\gamma ({\bf r})$, while the various chemical species $X_l$ each 
can have their own $T_l({\bf r}).$  Thus
\begin{eqnarray}
\Sigma_\gamma = \sum_l \int dV \int d\epsilon \int d\Omega\ I_\epsilon 
({\bf r},\epsilon ,\Omega )
 \Bigl[\frac{1}{T_\gamma({\bf r},\epsilon ,\Omega )}
- \frac{1}{T_l({\bf r})}\Bigr]\quad .
\end{eqnarray}
Again $I_\epsilon$ is the local specific radiation intensity emitted by
the matter.  Contributions such as (41) or (42) occur in addition to such 
gradient terms as (37) or (40).

In a supernova \cite{Smit}, the matter and the neutrinos can have different
temperatures $T$ and $T_\nu$, and $\Sigma_\nu$ picks up a contribution
analogous to (41).  The total $\Sigma$ again has other contributions, e.g.,
from gradients of $T$, $T_\gamma$, and $T_\nu$ such as (37) and (40).

Neutrinos can interact among themselves by weak neutral
currents and change their own phase space distribution without any ordinary 
matter present.  The associated entropy production is
\begin{eqnarray}
\Sigma^{\rm NC}_\nu & = & \int dV \int d\epsilon \int d\Omega 
\int d\epsilon^\prime \int d\Omega^\prime\ I_{\epsilon\epsilon^\prime}({\bf r},
\epsilon ,\epsilon^\prime , \Omega ,\Omega^\prime )\ \times \nonumber \\ 
 & & \Bigl[\frac{1}{T_\nu({\bf r},\epsilon ,\Omega )} -
 \frac{1}{T_\nu({\bf r}, \epsilon^\prime ,\Omega^\prime )}
\Bigr]~,
\end{eqnarray}
which is local in form.  $I_{\epsilon\epsilon^\prime}({\bf r},\epsilon,
\epsilon^\prime,\Omega,\Omega^\prime)$ is the local doubly-specific radiation
intensity of the neutrinos ``shining'' on themselves and is proportional to
the neutrino-neutrino weak neutral current reaction cross section.
$\Sigma^{\rm NC}_\nu$ 
vanishes if thermal equilibrium obtains and there is only a single temperature,
$T_{\bf p}$ = $T_{{\bf p^\prime}}$ for all ${\bf p},$ ${\bf p^\prime},$
at each point ${\bf r}.$

\section{Summary and Conclusion}

A non-zero density of entropy production, $\Sigma$, indicates a local
process of an irreversible nature. It is distinct from the local
density of entropy, and it has no direct implications for processes 
elsewhere. For any given process, one may measure local nearness
to equilibrium by its magnitude, although relative nearness of two
different processes may have no meaning by this measure.
If a single process is free, one
expects that the entropy production rate will relax toward zero. If some
exterior
conditions prevent this dynamical relaxation, the process will instead 
simply relax as far as the exterior arrangements permit. 

This dynamical argument anticipates minimum entropy production, at 
least
for sufficiently simple systems where thermodynamic quantities have 
meaning. But it also anticipates, as thermodynamics often does, an 
analogous static construction of the same principle, as the relaxation
may be imagined to be constrained to quasi-steady conditions in
whatever state it may be.

Thus the principle of minimum entropy
production found here is not unexpected. However in many respects it is
very different from the classical minimum entropy production principle
of Prigogine \cite{DeG}. It does not depend on Onsager's principle; 
there are no ``linear" empirical flux-force laws, nor is there even a
meaningful bilinear construction.

Under the LTE-NESS assumption, systems of both matter \cite{DeG}
and photons \cite{Essexa,Essexb,Essexc,KaB} exhibit some type of 
minimum entropy production.   A photon-like principle
holds, under the same assumptions, for neutrinos, as seen in examples
given in sections~IV and~V.  These examples can be generalised to many
local or continuously varying temperatures.

Conservation laws play identical roles in all three cases, by constraining
the microscopic interactions of the quanta. Thus, microscopic energy and
momentum are always conserved, with all the appropriate macroscopic 
consequences. Because neutrinos and photons
are both taken here as massless and not conserved in number, the 
analogy between these
two particles can be carried through in most aspects.  However, neutrinos
are fermions, which always have some associated conservation law; in this
case, lepton number $L.$  Because the weak interactions conserve $B - L,$
baryon number $B$ is also conserved.  Electromagnetic interactions conserve 
charge,
but since photons themselves do not carry charge, this conservation law
is trivial in radiative transfer, in contrast to the situation
for neutrinos, which do carry $L.$

The exact masslessness of neutrinos is not proven experimentally 
\cite{Gel}, and a
logical generalization of our results here is to extend the treatment to 
massive neutrinos.  Although we have used an electron chemical potential
$\mu_e$ in matter, another generalization left open is to include a 
neutrino chemical
potential $\mu_\nu$.  These two extensions will be presented in a subsequent
publication \cite{EaK}.

\acknowledgements

We thank the Telluride Summer Research Center, where part of this work 
was done. D.~C.~Kennedy acknowledges the support of the
University of Florida/Institute for Fundamental Theory, the U.S. 
Department of Energy under contract DE-FG05-86-ER40272 (U. Florida)
and the NASA/Fermilab Theoretical Astrophysics group under DOE/NASA contract
NAG5-2788.


\begin{references}

\bibitem{Sch}  
Schwarzschild,~K. 1906, {\it G\"{o}ttinger Nachrichten}, {\bf 195}, 41.

\bibitem{Ch}
Chandrasekhar,~S. 1950, {\it Radiative Transfer} (New York:
Dover Publications, 1960).

\bibitem{Essexa}
Essex,~C. 1984, {\it J. Planet. Space Sci.} {\bf 32}, 1035.

\bibitem{Essexb}
Essex,~C. 1984, {\it J. Atmos. Sci.} {\bf 41}, 1985.

\bibitem{DeG}
De~Groot~S.~R. and Mazur,~P. 1962, {\it 
Non-Equilibrium Thermodynamics} (New York: Dover Publications, 1984).

\bibitem{Essexad}
Essex,~C. 1990, in {\it Advances in Thermodynamics, Vol.~3:
Nonequilibrium Theory and Extremum Principles,} S.~Sieniutycz and
P.~Salamon, eds. (New York: Taylor and Francis) 435.

\bibitem{BoerGroot}
De~Boer,~W.~P.~H. and Van~Weert,~Ch.~G. 1977, {\it Physica} {87A}, 67, 80.
De~Groot,~S.~R., 1979, {\it Ann. Inst. Henri Poincar\' e} {\bf A31}, 377.

\bibitem{EaK}
Essex,~C. and Kennedy,~D.~C. 1998, in preparation.

\bibitem{Essexg}
Essex,~C. 1987,  {\it Geophys. Astrophys. Fluid Dynam.} {\bf38}, 1.

\bibitem{Essexc}  
Essex,~C. 1984, {\it Astrophys. J.} {\bf 285}, 279.

\bibitem{KaB} 
Kennedy,~D.~C. and Bludman,~S.~A. 1997, {\it Astrophys. J.} {\bf 484}, 329.

\bibitem{Les} 
Lesins,~G.~B. 1991, in {\it Scientists on Gaia,} S.~H.~Schneider
and P.~J.~Boston, eds. (Cambridge, Massachusetts: MIT Press) 121.

\bibitem{Smit} Smit,~J.~M. {\it et al.} 1996, {\it Astrophys. J.} {\bf 460}, 
895.

\bibitem{Gel}
R.~M.~Barnett {\it et al.} (Particle Data Group) 1996, {\it Phys. Rev.} 
{\bf D54}, 1, and 1998, World Wide Web URL http://pdg.lbl.gov/ .  

\end{references}
\end{document}